\documentclass{appolb}
\usepackage{graphicx}
\begin{document}
\title{Anisotropic critical fields of MgB$_2$ single crystals
\thanks{Presented at the Strongly Correlated Electron Systems
Conference, Krak\'ow 2002}%
}

% Authors and Affiliations

\author{T.~Dahm
\address{ Universit\"at T\"ubingen, Institut f\"ur Theoretische Physik,
Auf der Morgenstelle 14, 72076 T\"ubingen, Germany
  }
\and A.~I.~Posazhennikova
\address{ Laboratorium voor Vaste-Stoffisica en Magnetisme, Katholieke Universiteit Leuven,
 B-3001 Leuven, Belgium  }
\and
K.~Maki
\address{  Max-Planck-Institute for Physics of Complex Systems,
N\"othnitzer Str.~38, D-01187 Dresden, Germany \\
  Department of Physics and Astronomy, University of Southern
California, Los Angeles, CA 90089-0484, USA}
}
\maketitle

% Abstract

\begin{abstract}
The recently discovered superconductivity in MgB$_2$ has created the world
sensation. In spite of the relatively high superconducting transition
temperature $T_c=$~39~K, the superconductivity is understood in terms
of rare two gap superconductor with energy gaps attached to the
$\sigma$ and $\pi$-band. However, this simple model cannot describe
the temperature dependent anisotropy in $H_{c2}$ or the temperature
dependence of the anisotropic magnetic penetration depth. Here we
propose a model with two anisotropic energy gaps with different shapes.
Indeed the present model describes a number of pecularities of MgB$_2$
which have been revealed only recently through single crystal MgB$_2$.
\end{abstract}


\section{Introduction}

The discovery of new superconductivity in MgB$_2$ took the world
by surprise \cite{Akimitsu}. Early studies based on polycrystalline
samples lead to a two gap model \cite{Shulga,An,Liu}. On the other hand
the anisotropy in $H_{c2}(t,\Theta)$ suggested an anisotropic $s$-wave
model \cite{deLima,Angst,Chen,Haas,us}. Further it is clear that the
simple two gap model cannot describe the strong temperature dependence
of $\gamma(t)=H_{c2}^{ab}(t)/H_{c2}^c(t)$ observed in single crystal
MgB$_2$ \cite{Angst,us}. For this we need an order parameter
$\Delta(\vec{k})$ of oblate shape \cite{us}. On the other hand, it
is well known that $H_{c2}^{ab}(t)/H_{c2}^c(t)>1$ and
$H_{c1}^{ab}(t)/H_{c1}^c(t)>1$ for single crystal experiments
\cite{Buzea,Jin,Xu}, which contradicts the Ginzburg-Landau phenomenology.
Further, both magnetic penetration depth data and $H_{c1}^c(t)$ data
suggest a prolate order parameter as in \cite{Chen,Haas}. Indeed, an
earlier STM study suggested a prolate order parameter as well
\cite{Yeh}.

Is the order parameter prolate or oblate? The answer is that we need
both. We suggest, that the oblate order parameter is attached to
the $\sigma$-band while the prolate order parameter to the
$\pi$-band.
In the following we shall describe salient features of single
crystal MgB$_2$ within the present model.

\section{Upper critical field}

We just point out two features in the upper critical field which
are outside of the Ginzburg-Landau phenomenology.

(a) the strong temperature dependence of the anisotropy
parameter $\gamma(t)=H_{c2}^{ab}(t)/H_{c2}^c(t)$ \cite{us}.
(b) second the deviation from the effective mass model
$H_{c2}(t,\Theta)/H_{c2}(t,0)\neq (\cos^2\Theta + \gamma
\sin^2\Theta)^{-1/2}$ \cite{Angst}.

For example $\Delta(\vec{k}) \sim 1/\sqrt{1+a z^2}$, where $z=\cos
\Theta$ and $a\sim 100$ can describe the temperature dependence of
$\gamma(t)$, as has been shown in Ref. \cite{us}. Also, by fitting
the experimental data we obtained $v_a\simeq 2.7 \cdot10^7$~cm/sec
and $v_c/v_a\simeq 0.48$. We point out that these values are very
consistent with the ones for the $\sigma$-band \cite{Shulga}.

By fitting experimental data for $H_{c2}(\theta)$ \cite{Eltsev}
close to $T_c$ we can deduce the ratio $v_c/v_a \simeq 0.73$,
implying $a=10$, for example. This result is bigger than that
estimated in our previous analysis of $H_{c2}(t)$ \cite{us}. Next
order correction in $(T-T_c)/T_c$ leads to the increase of the
ratio $v_c/v_a$ for the same values of the parameter $a$. In the
future we are going to elaborate our study of the upper critical
field behavior in MgB$_2$ by taking into account impurity
scattering, what we believe will improve the agreement with
experiment.

\section{Lower
critical field}

\begin{figure}[!ht]
\begin{center}
\includegraphics[width=0.5\textwidth,angle=270]{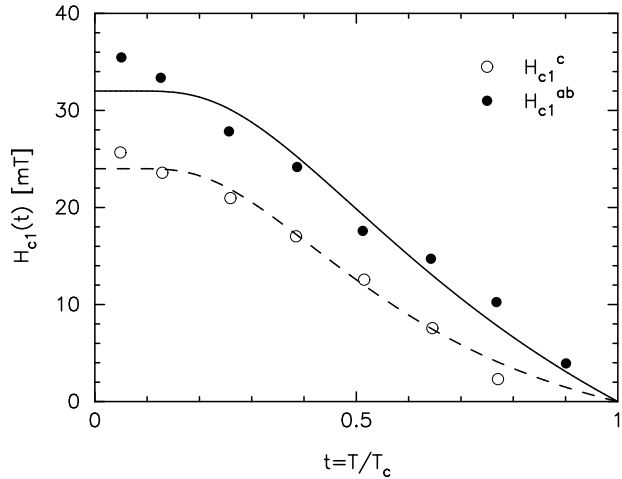}
\end{center}
\caption{Temperature dependence of the lower critical fields
$H_{c1}^c$ (dashed line) and $H_{c1}^{ab}$ (solid line) along
with the corresponding experimental data from Ref.\cite{Xu}. }
\label{fig2}
\end{figure}

>From $c$-axis oriented films and single crystals, the superfluid
density $\rho_{s,a}$ and the lower critical fields $H_{c1}^c$ and
$H_{c1}^{ab}$ have been extracted reliably \cite{Jin,Xu}. It is
clear this time that we need a prolate order parameter to fit
these data. We choose $\Delta(\vec{k})=1/\sqrt{1-a z^2}$ with
$a \simeq 0.92 \sim 0.95$. These values give a good fit of
penetration depth data \cite{Jin}. The magnetic penetration
depths are related to the lower critical fields via the formulas
\begin{eqnarray}
H_{c1}^c(t)&=&\frac{\Phi_0}{2 \pi \lambda_a^2(t)} \ln \frac{\lambda_a(t)}{
\xi_a(t)} \label{eq2} \\
H_{c1}^{ab}(t)&=&\frac{\Phi_0}{2 \pi \lambda_a(t) \lambda_c(t)}
\ln \sqrt{\frac{\lambda_a(t)\lambda_c(t)}{\xi_a(t)\xi_c(t)}} \label{eq3}
\end{eqnarray}
where $\lambda_a(t)$ and $\lambda_c(t)$ are the magnetic penetration
depth with the supercurrent in the ab-plane and in parallel to the c-axis,
respectively. They are related to the superfluid density via
$\rho_{s,a}(t)=\lambda_a^2(0)/\lambda_a^2(t)$ and
$\rho_{s,c}(t)=\lambda_c^2(0)/\lambda_c^2(t)$. Taking $a=0.95$, which
fits $\rho_{s,a}(t)$ from Ref. \cite{Jin}, we calculate
$\rho_{s,c}(t)$ and obtain the temperature dependences of
$H_{c1}^c$ and $H_{c1}^{ab}$ from Eqs. (\ref{eq2}) and (\ref{eq3}),
neglecting the temperature dependence of the logarithms. The
result is shown in Fig. \ref{fig2} along with the experimental
results from Ref. \cite{Xu}. From these fits we obtain
$H_{c1}^c(0)=24$~mT and $H_{c1}^{ab}(0)=32$~mT. From this the
ratio of the relevant Fermi velocities can be estimated as
$H_{c1}^{ab}(0)/H_{c1}^c(0) \approx v_c/v_a = 1.3$. In other
words, the corresponding Fermi surface is more isotropic and further
$v_c > v_a$. This strongly suggests that the prolate order parameter
we are considering has to be associated with the $\pi$-band.

\section{Synthese}

We have seen so far that we need two energy gaps of different
shape in order to describe $H_{c2}(t,\Theta)$  and
$H_{c1}(t,\Theta)$. The temperature dependence of $\gamma(t)$
indicates that an oblate order parameter
($\Delta(\vec{k})=1/\sqrt{1+a z^2}$) dominates the behavior at
high magnetic field. Also we need $v_c/v_a=0.48$. This suggests
the cylindrical Fermi surface associated with the $\sigma$-band as
the carrier of this order parameter. On the other hand, for
$H_{c1}(t,\Theta)$ we need a prolate order parameter
($\Delta(\vec{k})=1/\sqrt{1-a z^2}$). Also, the anisotropy in
$H_{c1}(t,\Theta)$ suggests the $\pi$-band as the carrier of this
order parameter. In other words, if we assume that the high field
properties are controlled by the oblate order parameter attached
to the $\sigma$-band, while the low field properties are due to
the prolate order parameter attached to the $\pi$-band, we have a
consistent picture for superconductivity in MgB$_2$.

We believe that this picture will be crucial to understand
also a variety of anomalies observed in the vortex state
in MgB$_2$.

We would like to thank M.~Angst, F.~Bouquet, S.~Haas, A.~Janossy,
A.~Junod, B.~B.~Jin and N.~Klein for useful discussions on MgB$_2$.


\begin{thebibliography}{11}

\bibitem{Akimitsu} J. Nagamatsu et al,
\textit{Nature(London)} \textbf{410}, 63 (2001).

\bibitem{Shulga} S. V. Shulga et al, cond-mat/0103154.

\bibitem{An} J.~M.~An and W.~E.~Pickett, \textit{Phys. Rev.
Lett.} \textbf{86}, 4366 (2001).

\bibitem{Liu} A. Y. Liu, I. I. Mazin, and J. Kortus, \textit{Phys. Rev.
Lett.} \textbf{87}, 087005 (2001).

\bibitem{deLima} O. F. de Lima et al., \textit{Phys. Rev. B} \textbf{64},
144517 (2001).

\bibitem{Angst} M. Angst et al,
\textit{Phys. Rev. Lett.} \textbf{88}, 167004 (2002).

\bibitem{Chen} Y. Chen, S. Haas, and K. Maki, \textit{Current Appl. Phys.}
\textbf{1}, 333 (2001).

\bibitem{Haas} S. Haas and K. Maki, \textit{Phys. Rev. B}
\textbf{65}, 020502(R) (2002).

\bibitem{us} A.~I.~Posazhennikova, T.~Dahm, and K.~Maki, cond-mat/0204272, to appear in
\textit{Europhys.Lett.
}

\bibitem{Buzea} C. Buzea and T. Yamashita, \textit{Supercond. Sci.
Technol.} \textbf{14}, R115 (2001).

\bibitem{Jin} B.~B.~Jin et al, cond-mat/0112350, to appear in \textit{Phys. Rev. B}.

\bibitem{Xu} M.~Xu et al, \textit{Appl. Phys. Lett.} \textbf{
79}, 2779 (2001).

\bibitem{Yeh} P.~Seneor et al, Phys. Rev. B {\bf 65}, 012505 (2002).

\bibitem{Eltsev} Yu.~Eltsev et al, Physica C (Proceedings of ISS2001, 
Kobe, Japan, Sept. 24-26, 2001), in press.

\end{thebibliography}
\end{document}